\documentclass[prstab,showpacs,preprintnumbers,amsmath,amssymb,showkeys]{revtex4}

\usepackage{graphicx}
\usepackage{bm}
\def\slashchar#1{\setbox0=\hbox{$#1$}
   \dimen0=\wd0 \setbox1=\hbox{/} \dimen1=\wd1
   \ifdim\dimen0>\dimen1 \rlap{\hbox to \dimen0{\hfil/\hfil}} #1
   \else  \rlap{\hbox to \dimen1{\hfil$#1$\hfil}} / \fi}

\begin{document}

\title{Strange particle production at low and intermediate energies}
\author{M. Rafi Alam$^*$}
\author{I. Ruiz Simo$^!$}
\author{M. Sajjad Athar$^*$}
\author{M. J. Vicente Vacas$^!$}
\affiliation{$^*$Department of Physics, Aligarh Muslim University, Aligarh-202 002, India}
\affiliation{$^!$Departamento de F\'\i sica Te\'orica and IFIC, Centro Mixto
Universidad de Valencia-CSIC, Institutos de Investigaci\'on de
Paterna, E-46071 Valencia, Spain}

\begin{abstract}
The  weak kaon production off the nucleon induced by neutrinos and antineutrinos is studied at low and intermediate energies
of interest for some ongoing and future neutrino oscillation experiments. We develop a microscopical model based on the SU(3) chiral Lagrangians. 
The studied mechanisms are the main source of kaon production for neutrino energies up to 2 GeV for 
the various channels and the cross sections are large enough to be amenable to be
 measured by experiments such as Minerva, T2K and NO$\nu$A.
\end{abstract}
\maketitle
\section{Introduction}
Neutrino physics has become one of the important areas of intense
 theoretical and experimental efforts. 
This is because neutrinos are instrumental in giving answer to some of the basic questions in cosmology, astro-, nuclear, and particle physics. 
The $\nu_\mu-N$ scattering processes have been studied by several authors 
for the quasielastic, 1 pion production and for deep-inelastic 
scattering processes~\cite{Boyd:2009}-\cite{Luis}. 
However, there are very few works where
 neutrino induced strange particle production have been been studied~\cite{Shrock:1975an}-\cite{Adera:2010zz}. 
These processes are important for the analysis of the 
precise determination of neutrino oscillation parameters.
In the few-GeV region it allows the detailed study of the strange-quark content of the
nucleon and gives some important information about 
the structure of the hadronic weak current.
Apart from that in the 
 atmospheric neutrino analysis these processes
gives $\Delta S$ backgrounds for
nucleon-decay searches. The antineutrino induced  $\Delta S=1$ single-hyperon
production can give us useful information about the
weak form-factors. A precise measurement of the 
hyperon cross-section specially $Q^2$ distribution 
will be useful in the determination of axial form 
factors. A better understanding of these processes will give more strength
to the basic understanding of V-A and Cabibbo theories.

Recently Miner$\nu$a is taking data in its first phase of
experiment with high statistics to explore
the strange physics. There is also probability of getting events of single kaons 
in the various beta-beam experiments in the energy region of ~1GeV. 
Also T2K is planning to run phase-II in 
 antineutrino mode and the NO$\nu$A experiments where the information about the single hyperon and single kaon
production may be obtained.

Strange particle
production via the weak interaction were initially studied by Shrock~\cite{Shrock:1975an}, Mecklenburg~\cite{Mecklenburg} and Dewan~\cite{Dewan}. Shrock~\cite{Shrock:1975an} and Mecklenburg~\cite{Mecklenburg}
independently studied the associated production of charged
current (CC) reactions by employing the Cabibbo theory with
SU(3) symmetry. Amer~\cite{Amer:1977fy} used harmonic oscillator quark model to estimate cross section for some of the associated production process. Dewan~\cite{Dewan} studied the CC and strangeness changing ($\Delta$S = 1) strange particle
production reactions. Recently Singh and Vicente Vacas~\cite{Singh:2006} have studied hyperon production induced by antineutrinos from nucleons and nuclei, Rafi Alam et al.~\cite{Rafi:2010} have studied single kaon production and 
 Adera et al.~\cite{Adera:2010zz} have studied differential cross section for $\nu$ induced C. C. Associated Particle Production. 

Most of the earlier neutrino experiments (1970's and '80s)  were performed using bubble chambers where cross-sections for many associated-production and $\Delta S = 1$ reactions were obtained using  
bubble chambers filled with Freon and/or Propane or with deuterium.  However, the data are statistically limited with large error bars\cite{Deden}-\cite{Mann_86}.

In this work, we have presented the results of our calculations for single kaon produced induced by neutrino/antineutrino reactions. In Sect.II, we present the formalism in brief and in Sect.III, the results and 
discussions are presented.
\section{Formalism}
The basic reaction for the $\nu(\bar\nu)$ induced charged current kaon production is
\begin{eqnarray}\label{reaction}
\nu_{l}(k) + N(p) \rightarrow l(k^{\prime}) + N^\prime(p^{\prime}) + K(p_{k}),\nonumber \\
\bar\nu_{l}(k) + N(p) \rightarrow l(k^{\prime}) + N^\prime(p^{\prime}) + \bar K(p_{k})
\end{eqnarray}
where $l=e,\mu$ and $ N \& N^\prime $=n,p.	
The expression for the differential cross section in lab frame for the above process is given by,
\begin{eqnarray}\label{d9_sigma}
d^{9}\sigma &=& \frac{1}{4 M E(2\pi)^{5}} \frac{d{\vec k}^{\prime}}{ (2 E_{l})} \frac{d{\vec p\,}^{\prime}}{ (2 E^{\prime}_{p})} \frac{d{\vec p}_{k}}{ (2 E_{K})} \delta^{4}(k+p-k^{\prime}-p^{\prime}-p_{k})\bar\Sigma\Sigma | \mathcal M |^2,
\end{eqnarray}
where  $ \vec{k}$ and $ \vec{k^\prime} $ are the 3-momenta of the incoming and outgoing leptons in the lab frame with energy $E$ and $ E^\prime$ respectively. The kaon lab momentum is $\vec{p}_k $ having energy $ E_K  $, $M$ is the nucleon mass,
$ \bar\Sigma\Sigma | \mathcal M |^2  $ is the square of the transition amplitude matrix element averaged(summed) over the spins of the initial(final) state. At low energies, this amplitude can be written in the usual form as
\begin{equation}
\label{eq:Gg}
 \mathcal M = \frac{G_F}{\sqrt{2}}\, j_\mu^{(L)} J^{\mu\,{(H)}}=\frac{g}{2\sqrt{2}}j_\mu^{(L)} \frac{1}{M_W^2}
\frac{g}{2\sqrt{2}}J^{\mu\,{(H)}},
\end{equation}
 where $j_\mu^{(L)}$ and $  J^{\mu\,(H)}$ are the leptonic and hadronic currents respectively, 
$G_F$ is the Fermi constant and 
$g$ is the gauge coupling.
The leptonic current can be readily obtained from the standard model Lagrangian coupling the $W$ bosons to the leptons 
\begin{equation}
{\cal L}=-\frac{g}{2\sqrt{2}}\left[{ W}^+_\mu\bar{\nu}_l
\gamma^\mu(1-\gamma_5)l+{ W}^-_\mu\bar{l}\gamma^\mu
(1-\gamma_5)\nu_l\right]
\end{equation}
\begin{figure}[t]
\includegraphics[height=.20\textheight,width=0.8\textwidth]{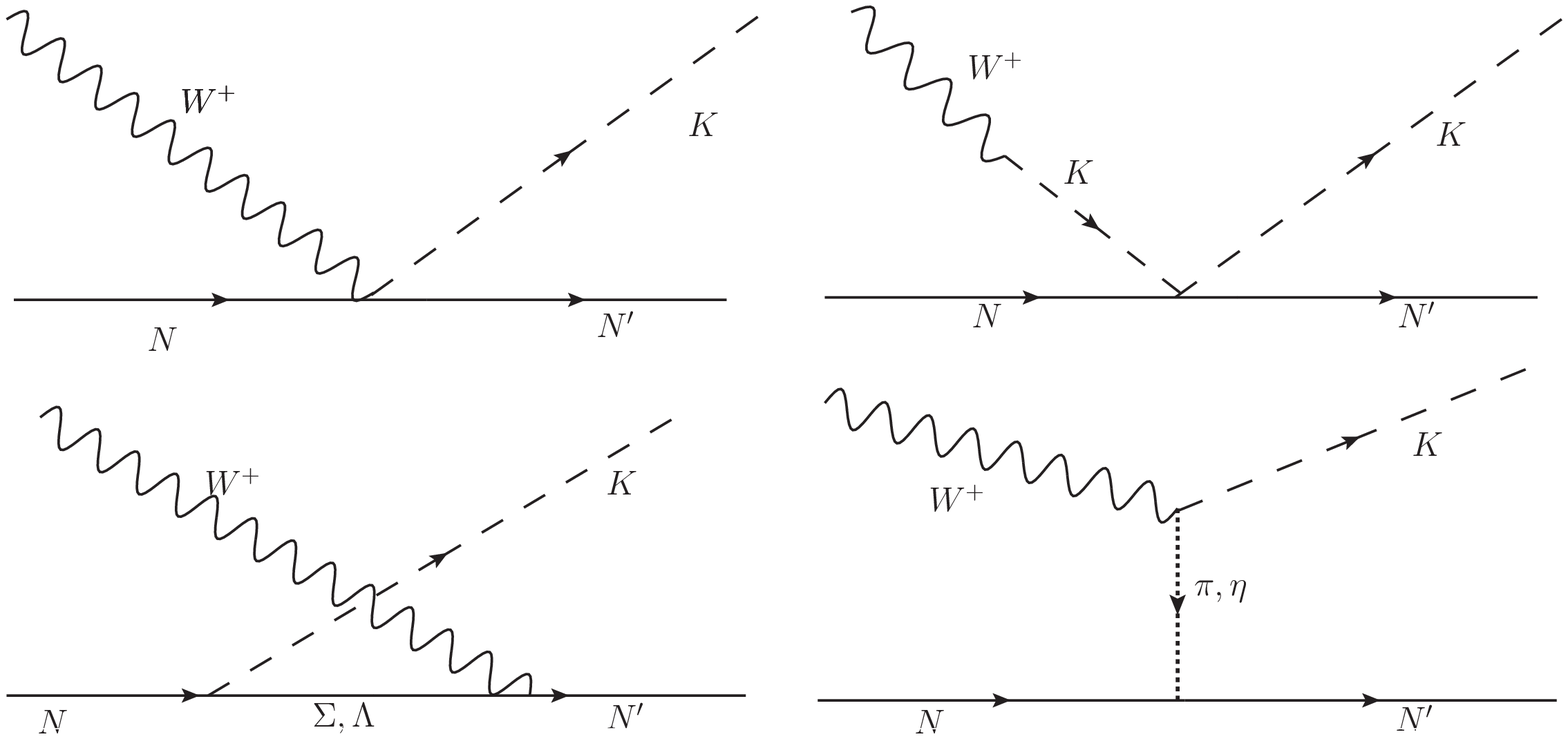}
\caption{Feynman diagrams for the process:$\nu_l N\rightarrow l^- N K$}
\label{fig:feynman_neutrino}
\end{figure}
\begin{figure}[t]
\includegraphics[height=.20\textheight,width=0.8\textwidth]{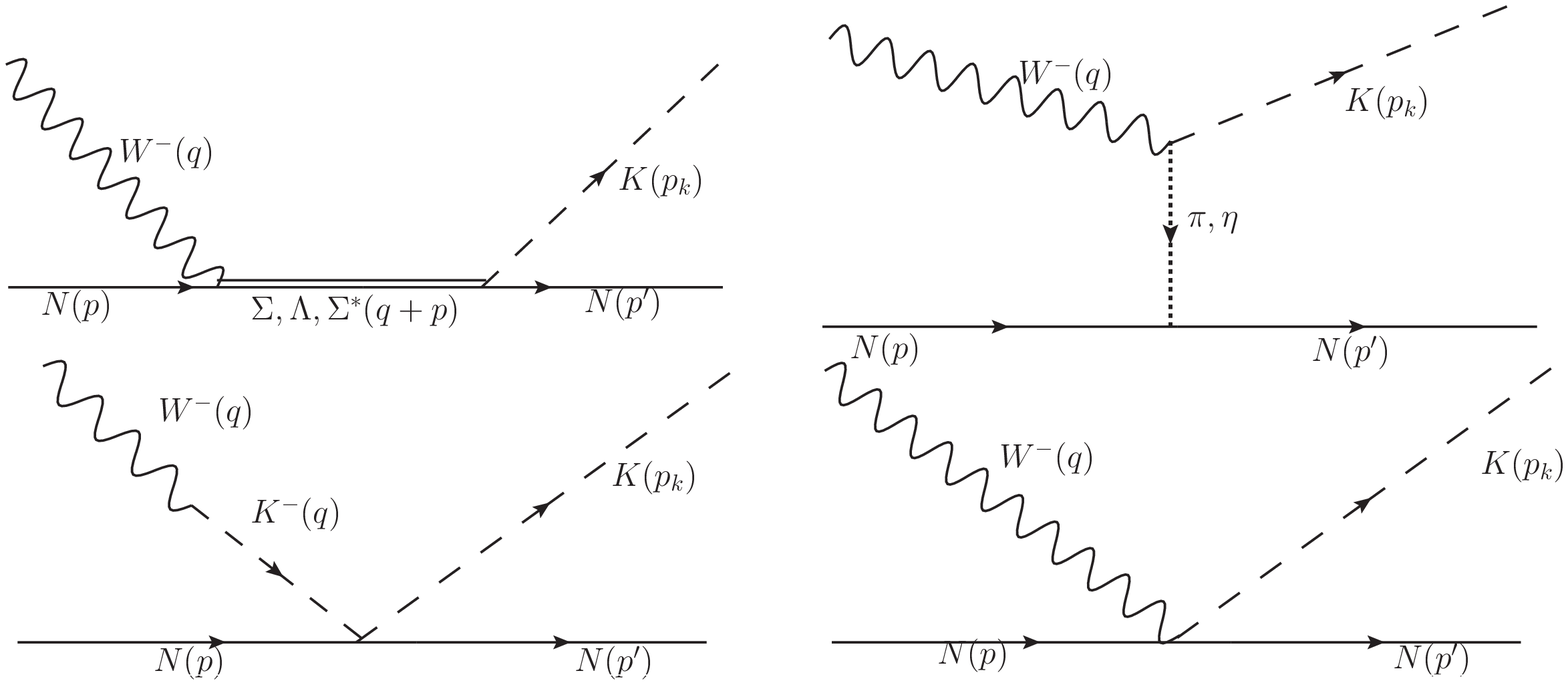}
\caption{Feynman diagrams for the process $\bar \nu_\mu n \rightarrow \mu^+ n K^-$}
\label{fig:feynman_antineutrino}
\end{figure}
In the case of neutrino induced kaon production process, we have considered four different channels that contribute to the hadronic current. 
They are depicted in Fig.~\ref{fig:feynman_neutrino}. There is a contact term (CT), a kaon pole (KP) term, a u-channel process with a $\Sigma$
or $\Lambda$ hyperon in the intermediate state and finally a meson ($\pi,\,\eta$) exchange term. For the specific reactions under consideration, 
there is no s-channel contributions given the absence of $S=1$ baryonic resonances. KP term is  proportional lepton mass and therefore its contribution is very small. 
While in the case of antineutrino induced kaon production process, besides the processes mentioned for the neutrino case, there are contributions 
from s-channel $\Sigma,\Lambda $ propagator, s-channel $\Sigma^*$
 resonance terms(Fig.~\ref{fig:feynman_antineutrino}).

The contribution of the different terms can be obtained in a systematic manner using Chiral Perturbation Theory ($\chi$PT). 
The lowest-order SU(3) chiral Lagrangian describing the pseudoscalar mesons in the presence of an external current is~\cite{Rafi:2010}:
\begin{equation}
\label{eq:lagM}
{\cal L}_M^{(2)}=\frac{f_\pi^2}{4}\mbox{Tr}[D_\mu U (D^\mu U)^\dagger]
+\frac{f_\pi^2}{4}\mbox{Tr}(\chi U^\dagger + U\chi^\dagger),
\end{equation}
where the parameter $f_\pi=92.4$MeV is the pion  decay constant, $U$ is the SU(3) representation of the meson fields~\cite{Rafi:2010} 
and $D_\mu U$ is its covariant derivative.
\begin{figure}[b]
\includegraphics[height=.25\textheight,width=0.5\textwidth]{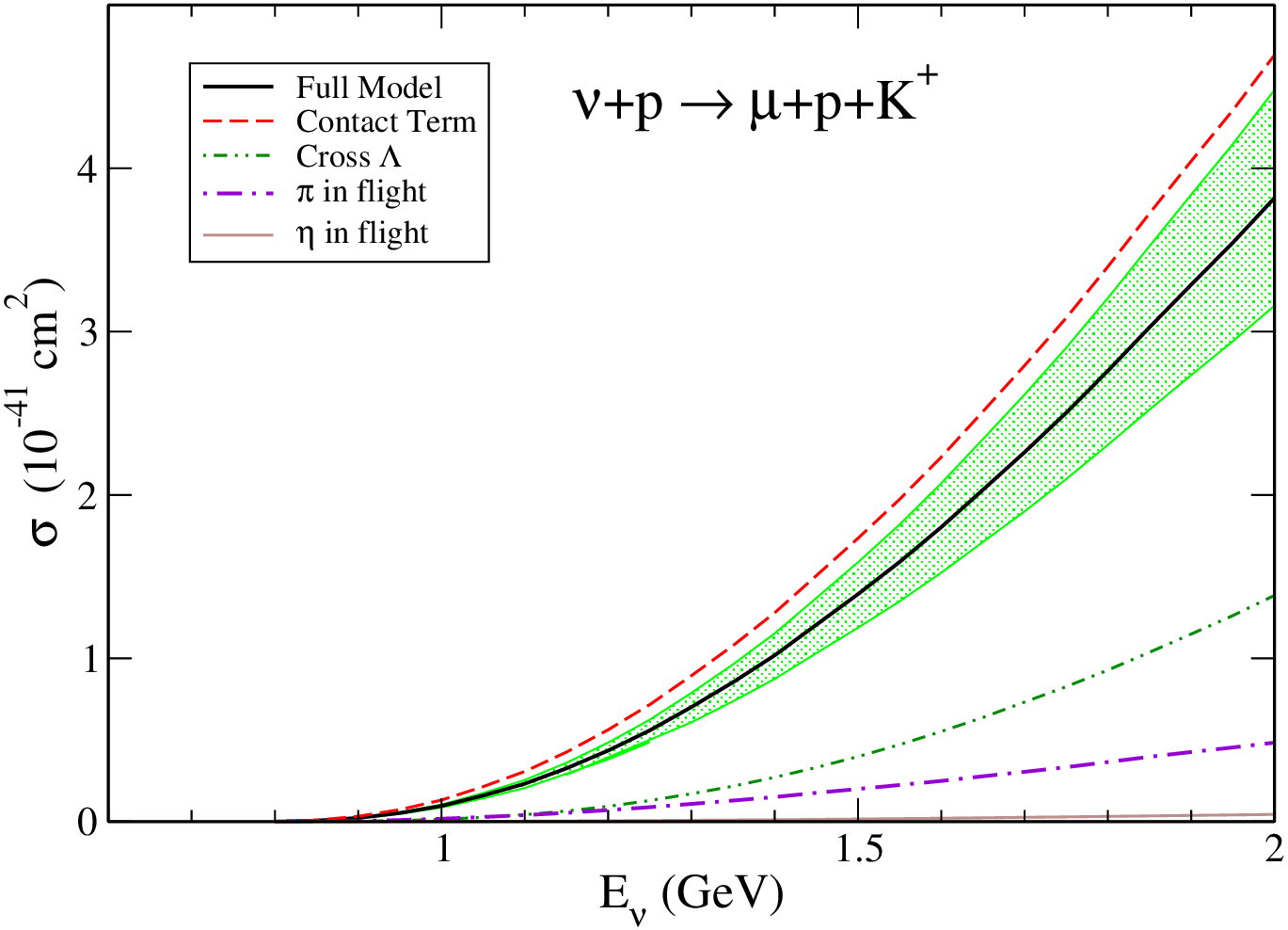}
\includegraphics[height=.25\textheight,width=0.5\textwidth]{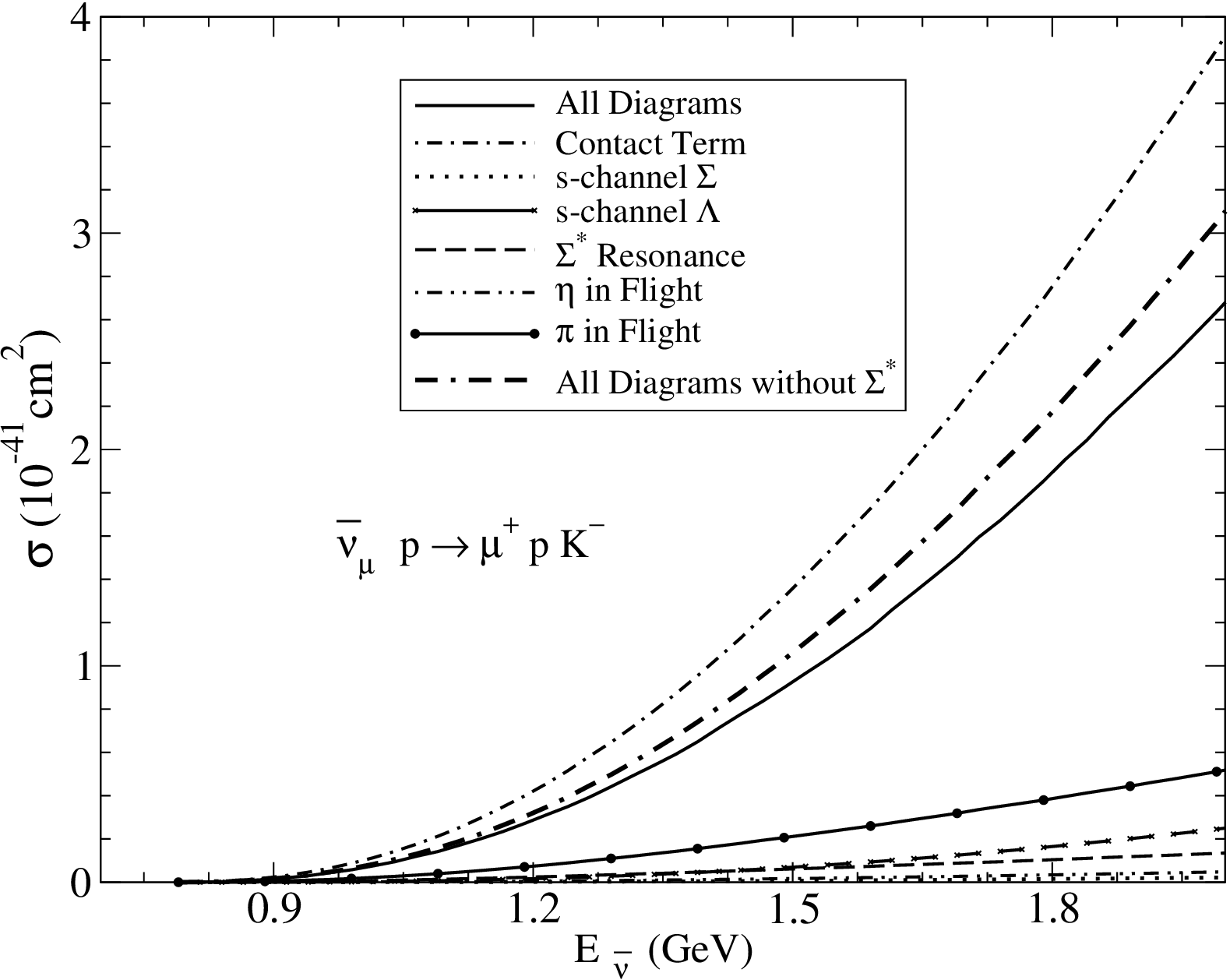}
\caption{Cross section for (Left panel) $\nu_\mu p \rightarrow \mu^- p K^+$ and (Right panel) $\bar \nu_\mu p \rightarrow \mu^+ p K^-$}
\label{fig:xsec_pp_nu}
\end{figure}
The lowest-order  chiral Lagrangian for the baryon octet in the presence of an external current can be written in terms of the Baryon SU(3) matrix as~\cite{Rafi:2010}:
\begin{equation}
\label{eq:lagB}
{\cal L}_{MB}^{(1)}=\mbox{Tr}\left[ \bar B  \left( i \rm{D}
-M \right) B \right]
-\frac{D}{2} \mbox{Tr}\left( \bar B \gamma^\mu \gamma_5 \{ u_\mu,B \} \right)
-\frac{F}{2}\mbox{Tr}\left( \bar B \gamma^\mu\gamma_5 \left[ u_\mu,B \right] \right),
\end{equation}
where $M$ denotes the mass of the baryon octet, B is the Baryon SU(3) matrix and the parameters $D=0.804$ and $F=0.463$
which are determined from the semileptonic decays.
\begin{figure}
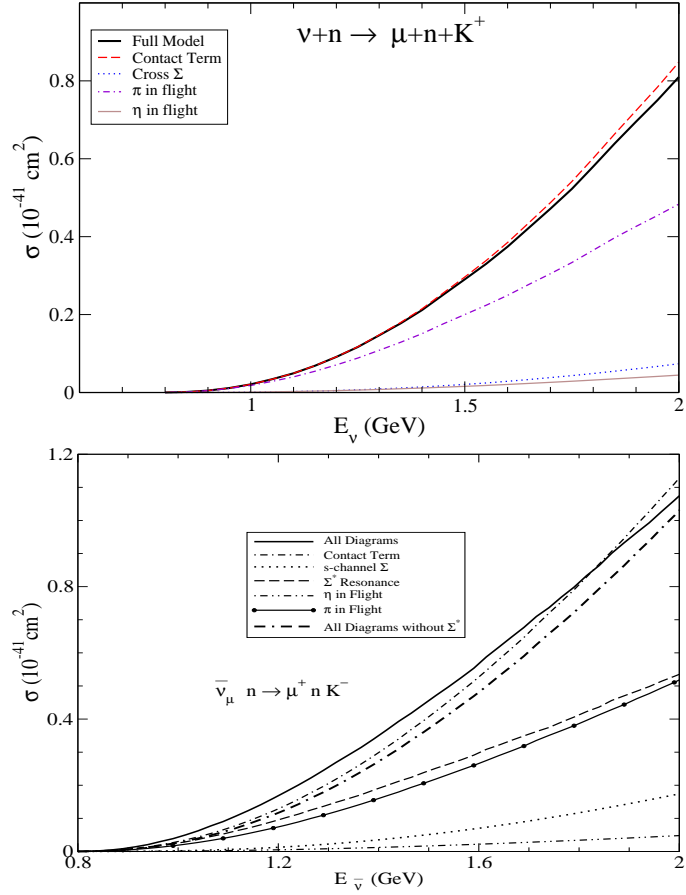

\includegraphics[height=.25\textheight,width=0.5\textwidth]{NN_nu.eps}
\includegraphics[height=.25\textheight,width=0.5\textwidth]{NN_muon_linear_v2.eps}
\caption{Cross section for (Left panel) $\nu_\mu n \rightarrow \mu^- n K^+$ and (Right panel) $\bar \nu_\mu n \rightarrow \mu^+ n K^-$}
\label{fig:xsec_nn_nu}
\end{figure}
At intermediate energies, we found that the weak excitation of the 
$\Sigma^*(1385)$ resonances and its subsequent decay in $NK$ is also 
important. To calculate amplitudes associated with $\Sigma^*$ we first
parameterize the $W^- N \rightarrow \Sigma^*$.
For this, we can write the most general form of the vector and axial-vector matrix element as,
\begin{eqnarray*} \label{eq:delta_amp}
\langle\Sigma^{*}; P= p+q\, | V^\mu | N;
p \rangle&=& V_{us} \bar u_\alpha(\vec{P} ) \Gamma^{\alpha\mu}_V \left(p,q \right)
u(\vec{p}\,), \\
\langle\Sigma^{*}; P= p+q\, | A^\mu | N;
p \rangle &=& V_{us} \bar u_\alpha(\vec{p} ) \Gamma^{\alpha\mu}_A \left(p,q \right)
u(\vec{p}\,)
\end{eqnarray*}
where
\begin{eqnarray}
\Gamma^{\alpha\mu}_V (p,q) &=&
\left [ \frac{C_3^V}{M}\left(g^{\alpha\mu} \slashchar{q}-
q^\alpha\gamma^\mu\right) + \frac{C_4^V}{M^2} \left(g^{\alpha\mu}
q\cdot P- q^\alpha P^\mu\right)
+ \frac{C_5^V}{M^2} \left(g^{\alpha\mu}
q\cdot p- q^\alpha p^\mu\right) + C_6^V g^{\mu\alpha}
\right ]\gamma_5 \nonumber\\
\Gamma^{\alpha\mu}_A (p,q) &=& \left [ \frac{C_3^A}{M}\left(g^{\alpha\mu} \slashchar{q}-
q^\alpha\gamma^\mu\right) + \frac{C^A_4}{M^2} \left(g^{\alpha\mu}
q\cdot P- q^\alpha P^\mu\right)
+ C_5^A g^{\alpha\mu} + \frac{C_6^A}{M^2} q^\mu q^\alpha
\right ]. \label{eq:del_ffs}
 \end{eqnarray}
In the above expression $C^{V,A}_{3,4,5,6}$ are the  $q^2$ dependent scalar and real vector
and axial vector form factors and 
$u_\alpha $ is the Rarita-Schwinger spinor. It is the $C_5^A$ term which is most dominant and we have considered only the terms with $C_5^A$ in the present work. 

The spin 3/2  propagator in the momentum space is given by,
\begin{equation}
G^{\mu\nu}(P)= \frac{P^{\mu\nu}_{RS}(P)}{P^2-M_{\Sigma^*}^2+ i M_{\Sigma^*} \Gamma_{\Sigma^*}},
\end{equation}
where $P^{\mu \nu}_{RS}$ is the spin 3/2 Rarita-Schwinger projection operator and $M_{\Sigma^*}$ is the resonance mass ($\sim 1385 MeV$). The $\Sigma^*$ decay width $\Gamma$ is around 
36$\pm$5MeV, however, we have taken P-wave decay width which is given as 
\begin{equation}
 \Gamma_{\Sigma^*}(W)=\frac{1}{192\pi}\left(\frac{\cal C}{f_\pi}\right)^2
\frac{\left((W+M)^2-m^2_\pi\right)}{W^5}\lambda^{3/2}(W^2,M^2,m^2_\pi) \;
\Theta(W-M-m_\pi)
\end{equation}
where $\lambda(x,y,z)=(x-y-z)^2-4yz$ is Callen lambda function and $\Theta$ is the 
step function. ${\cal C}$ is the KN$\Sigma^*$ coupling strength taken to be 1 in the present calculation.
\begin{figure}
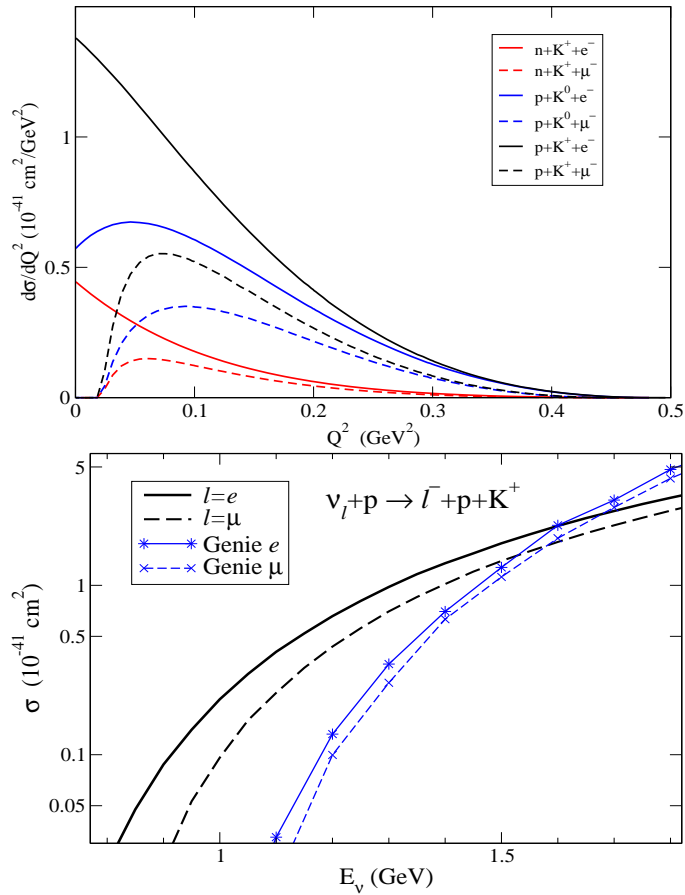

\includegraphics[height=.25\textheight,width=0.5\textwidth]{dsq2.eps}
\includegraphics[height=.25\textheight,width=0.5\textwidth]{genie.eps}
\caption{(Left panel) $\frac{d \sigma}{ d Q^2} $ at $ E_\nu=1 GeV $ for single kaon production induced by neutrinos. The curves 
are labeled according to the final state of the process. (Right panel) Cross sections as a function of the neutrino energy for single kaon production vs. associated production obtained with Genie~\cite{Andreopoulos:2009rq}. }
\label{fig:xsec_dq2_genie}
\end{figure}
\section{Results and Discussion}
The total scattering cross section $\sigma$ has been obtained by using Eq.~(\ref{d9_sigma}) after integrating over the kinematical variables. In the left panel of Fig.~(\ref{fig:xsec_pp_nu}), 
we present the results of the contributions of the different diagrams to the total cross sections for the neutrino induced process. The kaon pole contributions are negligible at the studied energies and are not shown
 in the figures although they are included in the full model curves. We observe the relevance of the contact term, not included in previous calculations. We find that the contact term is in fact dominant, followed by the u-channel diagram with a $\Lambda$ intermediate state and
 the $\pi$ exchange term. As observed by Dewan~\cite{Dewan} the u-channel $\Sigma$ contribution is much less important, basically because of the larger coupling  ($N K\Lambda \gg N K \Sigma$) of the strong vertex.
The curve labeled as Full Model has been calculated with a dipole form factor with a mass of 1 GeV. The band corresponds to changing up and down this mass by a 10 percent.
On the right panel of Fig.~(\ref{fig:xsec_pp_nu}), we have presented the results for antineutrinos: $\bar \nu_\mu + p\rightarrow \mu^+ + p + K^- $. We find that the contact term is the most 
dominant one followed by pion in flight and 
the s-channel diagram with $\Sigma^*$-resonance and $\Lambda$ as the 
intermediate states. The suppression of $\Sigma$ as the intermediate state is due to the difference in the coupling 
strength $g_{NK\Lambda} >> g_{NK\Sigma}$ and in the $\eta$ in flight due to $m_\eta > m_\pi$.
We also checked the effects of the $\Sigma^*(P_{13})$ resonance at the said energies.
We find that unlike the $\Delta (P_{33})$ dominance in 
pion production the contribution of $\Sigma^*$ is not too large. In Fig.~(\ref{fig:xsec_nn_nu}), corresponding results for $\nu_\mu + n \rightarrow  \mu^- + K^+ + n$ and 
$\bar\nu_\mu + n \rightarrow  \mu^+ + K^- + n$ processes are shown. In Fig.~(\ref{fig:xsec_dq2_genie}), we have shown the results for the $Q^2$ distribution for the three studied channels 
at a neutrino energy E$_\nu= 1$ GeV. 
The reactions are always forward peaked (for the final lepton), even in the absence of any form factor ($F(q^2)=1$), favoring relatively small values of the momentum transfer. 
On the right panel of the  Fig.~(\ref{fig:xsec_dq2_genie}), we have compared our results for the $\nu_\mu + p \rightarrow  \mu^- + K^+ + p$ process with the values
 for the associated production obtained by means of the GENIE Monte Carlo program~\cite{Andreopoulos:2009rq}.
We observe that, due to the difference between the energy thresholds, single kaon production for the
$\nu_l + p \rightarrow  l^- + K^+ + p$ is  clearly dominant for neutrinos of energies below 1.5 GeV. For the other two channels associated production becomes comparable at lower energies. Still,
single $K^0$ production off neutrons is larger than the associated production up to 1.3 GeV and even the much smaller $K^+$ production off neutrons is larger than the associated production up to 1.1 GeV.
The consideration of these $\Delta S=1$ channels is therefore important for the description of strangeness production for all low energy neutrino 
spectra and should be incorporated in the experimental analysis.

\end{document}